\documentstyle[preprint,aps,floats,epsfig]{revtex}

\begin{document}

\draft

\preprint{\rightline{ANL-HEP-PR-00-049}}

\title{Tricritical scaling at the $N_t=6$ chiral phase transition for 
       2~flavour lattice QCD with staggered quarks.}

\author{J.~B.~Kogut}
\address{Dept. of Physics, University of Illinois, 1110 West Green Street,
Urbana, IL 61801-3080, USA}
\author{D.~K.~Sinclair}
\address{HEP Division, Argonne National Laboratory, 9700 South Cass Avenue,
Argonne, IL 60439, USA}

\maketitle

\begin{abstract}
We have simulated lattice QCD directly in the chiral limit of zero quark mass
by adding an additional, irrelevant 4-fermion interaction to the standard
action. Using lattices having temporal extent of six and spatial extents of
twelve and eighteen, we find that the theory with 2 massless staggered quark
flavors has a second order finite temperature phase transition. The critical
exponents $\beta_{mag}$, $\delta$ and $\nu$ are measured and favour tricritical
behaviour over that expected by universality arguments. The pion screening
mass is consistent with zero below the transition, but is degenerate with the
nonzero $\sigma(f_0)$ mass above the transition, indicating the restoration of
chiral symmetry.
\end{abstract}


\pagestyle{plain}
\parskip 5pt
\parindent 0.5in

\narrowtext

There are several bottlenecks in the simulation studies of lattice field
theories by standard algorithms \cite{HMC}, \cite{HMD}. One of the most
harmful is that the chiral limit cannot be simulated directly because the
standard algorithms fail to converge in the limit of vanishing bare fermion
mass and become prohibitively expensive at small quark masses.
Various chiral transitions, such as the finite temperature Quantum
Chromodynamics (QCD) transition to be discussed here, have been difficult to
study quantitatively because of this. In particular, studies of critical
scaling and the extraction of critical exponents have been especially
difficult. This has led people to extract the critical exponents associated
with the mass dependence of the heights and positions of susceptibility peaks
\cite{karsch,aoki} and to compare the scaling behaviour of the chiral
condensate as both the coupling constant and the quark mass are varied
\cite{milc}, rather than studying the scaling of order parameters in terms of
the coupling constant and mass independently. The most extensive data has been
obtained for $N_t=4$ where it differs significantly from what universality
arguments predict.

These problems are largely solved by adding a four-fermi interaction to
standard lattice actions with staggered fermions \cite{LKS}.  The resulting
lattice action can be simulated directly in the chiral limit because an
auxiliary scalar field $\sigma$ (essentially the chiral condensate
$\langle\bar\psi\psi\rangle$) acts as a dynamical (`constituent') quark mass
term insuring that the inversion of the Dirac operator will be successful and
very fast. This approach is far more efficient and physical than the
traditional lattice algorithm in which the chiral limit is singular and a
nonzero bare quark mass $m$ is a necessity.

Using this strategy we have simulated lattice Quantum Chromodynamics (QCD) with
two flavors of staggered quarks on $12^3 \times 6$ and $18^3 \times 6$
lattices, at zero quark mass, in order to determine the position and nature of
the finite temperature transition. In addition to measuring the standard order
parameters, we have measured the pion and $\sigma$($f_0$) screening masses to
probe the nature of chiral symmetry restoration at this transition. We find
that the two flavor theory experiences a second order chiral symmetry
restoring phase transition. We make independent determinations of the critical
indices $\beta_{mag}$, $\nu$, and $\delta$. $\beta_{mag}$
differs significantly from that of the $O(4)$ (or $O(2)$) sigma model in 3
dimensions, expected in the standard scenario of dimensional reduction and
universality for this transition \cite{pisarski}, but is in excellent agreement
with that of the 3-dimensonal tricritical point. Since we have found that the
$N_t=4$ transition is first order at $\gamma=10$ \cite{LKS} and a tricritical
point would be expected to separate a line of first order transitions from a
line of second order transitions, this suggests that this theory would have a
normal critical point with the critical exponents expected from universality
for $N_t \ge 8$. If so, the anomalous behaviour of the $N_t=4,6$ transitions
is due to irrelevant terms in the lattice action which are important only on
coarse lattices.

In this letter we review the `standard' expectations and alternative scenarios
for the finite temperature transition in continuum and lattice QCD.  We review
the lattice formulation of QCD with chiral four-fermion interactions, and
present our data and fits. We end with discussions and conclusions, and
directions for future research.

Dimensional reduction and universality predict that 2-flavour QCD has the
critical behaviour of the 3-dimensional $O(4)$ sigma model \cite{pisarski}.
This theory has a second order phase transition with critical indices
\cite{Nickel}, $\nu = .73(2)$, $\beta_{mag} = .38(1)$, $\delta = 4.82(5)$,
etc. The lack of full flavour symmetry of the staggered fermion method is
expected to reduce this to an $O(2)$ sigma model with critical indices $\nu =
.71(2)$, $\beta_{mag} = .35(1)$ and $\delta = 4.81(1)$ \cite{Nickel}. A
breakdown of such arguments or the presence of lattice artifacts can produce
a first order transition. If both a first and simple second order transitions
are possible, so is a tricritical point for which $\nu=\frac{1}{2}$,
$\beta_{mag}=\frac{1}{4}$ and $\delta=5$ or $3$, the second value being for a
second independent symmetry breaking operator this more complex transition
admits \cite{dl}. Finally, if dimensional reduction fails to occur, one would
expect a mean field transition with $\nu=\frac{1}{2}$,
$\beta_{mag}=\frac{1}{2}$ and $\delta=3$.

This suggests that a direct measurement of $\beta_{mag}$, requiring simulations
at $m=0$ will be crucial in determining the nature of the finite temperature 
chiral transition in 2-flavour (lattice) QCD.

The lattice QCD action used in this study contains an additional chirally
invariant 4-fermion interaction. Because the extra interaction is irrelevant,
such an action should lie in the same universality class as the standard
action. Ideally, such an interaction should be chosen to have the $SU(N_f)
\times SU(N_f)$ flavour symmetry of the original QCD action. However, when one
introduces auxiliary scalar and pseudoscalar fields to render this action
quadratic in the fermion fields --- which is necessary for lattice
simulations ---, the fermion determinant is no longer real, even in the
continuum limit. Thus for 2-flavour QCD ($N_f=2$), we make a simpler choice and
choose a 4-fermion term with the symmetry $U(1) \times U(1) \subset SU(2)
\times SU(2)$, where $U(1) \times U(1)$ is generated by
$(\tau_3,\gamma_5\tau_3)$, which preserves the symmetries of the lattice action.
The euclidean Lagrangian density for this theory is
\begin{equation}
{\cal L}=\frac{1}{4}F_{\mu\nu}F_{\mu\nu}
        +\bar{\psi}(D\!\!\!\!/+\sigma+i\pi\gamma_5\tau_3+m)\psi
        + {   \gamma N_f \over 2}(\sigma^2+\pi^2)
\label{eqn:lagrangian1}
\end{equation}
when we have introduced auxiliary fields $\sigma$ and $\pi$ to render it
quadratic in the fermion fields. $\gamma \equiv 3/\lambda^2$, where the quartic 
term in the original Lagrangian had coefficient $\lambda^2/6 N_f$. The
molecular dynamics Lagrangian for a particular staggered fermion lattice
transcription of this theory has been given in reference \cite{LKS}. There we
reported simulation results on lattices with $N_t=4$ and discussed the choice
of parameters as well as systematic errors.  Here we shall discuss higher
precision simulations on larger lattices, $18^3 \times 6$ and $12^3 \times 6$.

By doing simulations on two lattice sizes we were able to isolate finite volume
effects in several observables and check that they had the behaviour expected
of finite size scaling theory. We ran simulations at $\gamma = 10$ and $20$ on
a $12^3 \times6$ and $\gamma = 20$ on an $18^3 \times 6$ lattice. We carefully
checked that our results do not depend on $\gamma$ or on the lattice size.
These technically important results will be discussed at greater
length elsewhere \cite{PRO}.

Since our $m=0$ lattice action does not single out a preferred chiral
direction, we accumulate the magnitude of the order parameter,
$\sqrt{\langle\bar{\psi}\psi\rangle^2+
\langle\bar{\psi}\gamma_5\tau_3\psi\rangle^2}$, 
rather than the order parameter itself, which averages to zero. This is not a
true order parameter since it does not vanish identically in the symmetric
phase. However, for a large system of volume $V$, its values in the high
temperature phase are of order $1/ \sqrt{V}$. In the low temperature phase the
order parameter can be fit to a standard critical form, $A(\beta_c -
\beta)^{\beta_{mag}}$, if $\beta$ is chosen within the model's scaling window
near $\beta_c$. Of course, we cannot simulate the model with $\beta$ chosen too
close to $\beta_c$ where the order parameter is very small, and finite size
rounding in this effective order parameter is considerable. This forces us to
work slightly away from the critical point and would produce indecisive
results if the scaling window were particularly small. These potential
problems bear close scrutiny.

The simulations of the lattice version of Equation~\ref{eqn:lagrangian1} were
performed using the hybrid molecular-dynamics algorithm with ``noisy''
fermions allowing us to tune $N_f$ to 2 flavours \cite{LKS}. The simulations
on an $18^3 \times 6$ lattice were performed at $14$ $\beta$ values from
$5.39$ to $5.45$, the observed chiral transition occurring close to
$\beta=5.423$. This enabled us to fit order parameters in the scaling window
on both sides of the transition. Runs at single $\beta$'s varied in length
from 5,000 to 50,000 molecular-dynamics time units close to the critical point
(this computer ``time'' is the same as that used in related works by other
groups, such as \cite{htmcgc8}). We were forced to make such lengthy runs
because of critical slowing down. Order parameters measured included the local
chiral condensates $\langle\bar{\psi}\psi\rangle$ and
$\langle\bar{\psi}\gamma_5\tau_3\psi\rangle$, their counterparts in terms of
the auxiliary fields viz $\langle\sigma\rangle$ and $\langle\pi\rangle$, the
thermal Wilson line, the plaquette, and operators contributing to the partial
entropies of each field. These were measured every 2 time units and binned to
minimize the effect of correlations on error estimates. The molecular dynamics
``time'' increment for updating was $dt = 0.05$ \cite{LKS}. This was large
enough that the order parameters $\langle\sigma\rangle$ and
$\langle\pi\rangle$ have sizable ${\cal O}(dt^2)$ errors. However, since these
errors can be absorbed in a redefinition of the couplings in the action, they
should not affect universality.

\begin{figure}[htb]
\epsfxsize=6in
\centerline{\epsffile{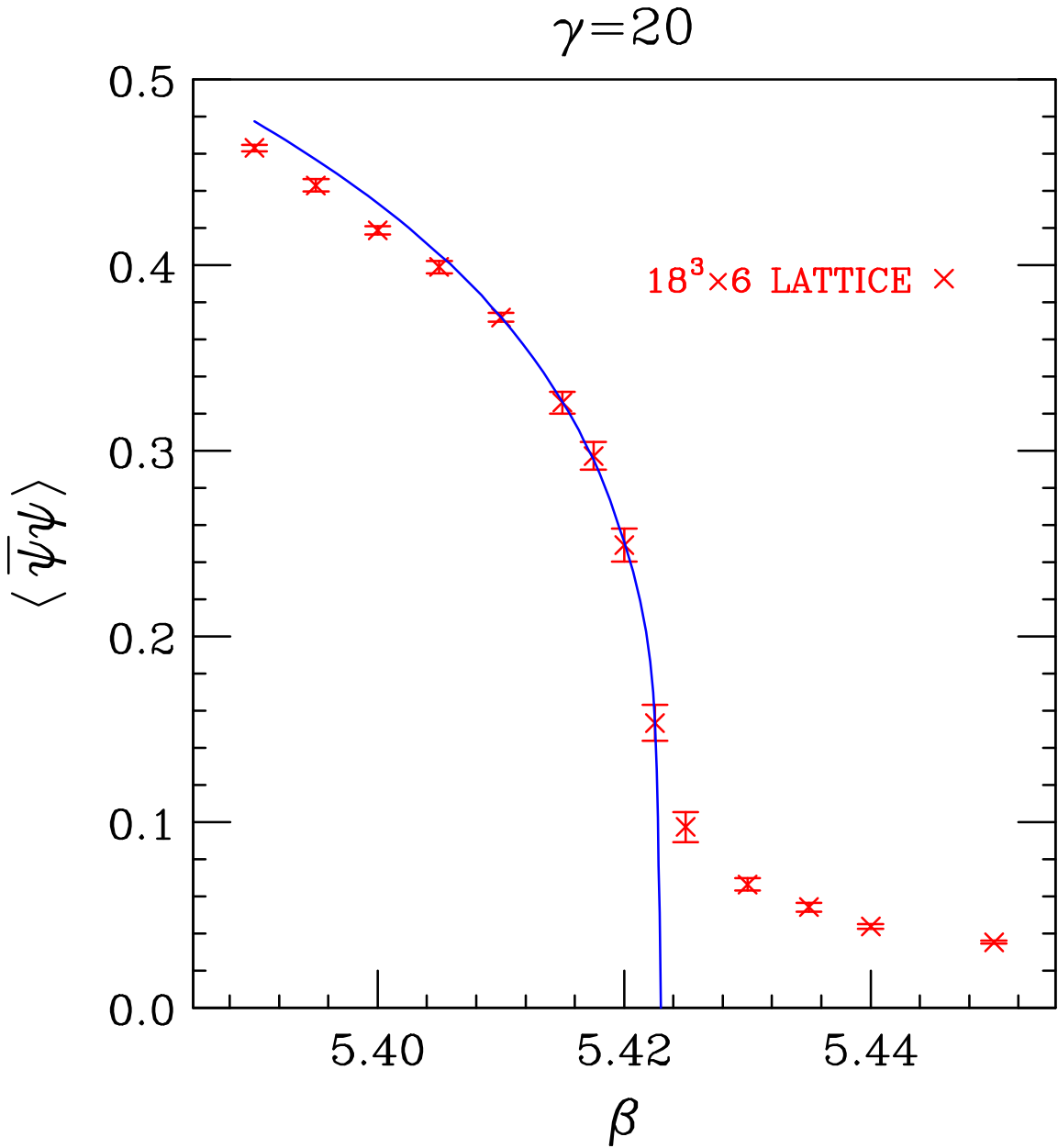}}
\caption{The order parameter plotted against $\beta$. The solid line is the 
fit discussed in the text.\label{fig:mag618}}
\end{figure}

The data for the magnitude of the order parameter
$\sqrt{\langle\bar{\psi}\psi\rangle^2+
\langle\bar{\psi}\gamma_5\tau_3\psi\rangle^2}$ is plotted against $\beta$ in
Figure~\ref{fig:mag618}. The points between $\beta$ values of $5.41$ and
$5.4225$ were fitted to a simple powerlaw $A(\beta_c - \beta)^{\beta_{mag}}$.
The fit has a confidence level of 94 percent and determined the parameters
$\beta_c = 5.4230(2)$, $a = 1.19(12)$ and, most importantly, the critical index
$\beta_{mag} = 0.27(2)$, in good agreement with the $0.25$ of the tricritical
point, but inconsistent with the $O(4)$, $O(2)$, and mean field values.

We see in the figure that points at stronger coupling than $\beta=5.41$ are
outside the scaling window. For example, if the points at $\beta = 5.4$ and
$\beta = 5.405$ were included in the fit, the confidence level would fall to
28 percent while the predicted best value for $\beta_{mag}$ does not change
significantly. Our simulations on $12^3 \times 6$ lattices produced values for
the order parameter which were essentially identical to those shown in the
figure for $\beta \leq 5.4225$, but were considerably larger for $\beta >
5.4225$. Using the 2 lattice sizes to remove the finite volume effects gave
$\beta_{mag}=0.24(2)$ consistent with the above. The data on the weak coupling
side of the transition is, in fact, compatible with the finite size scaling
prediction that the magnitude of the order parameter should fall to zero as
$1/\sqrt{V_s}$, where $V_s$ is the volume of the spatial box. For the $12^3
\times 6$ lattice and $\gamma = 10$, our fits predicted $\beta_{mag} =
0.27(3)$, in good agreement with the $18^3 \times 6$, $\gamma=20$ results,
while $\beta_c = 5.4650(1)$.

\begin{figure}[htb] 
\epsfxsize=6in
\centerline{\epsffile{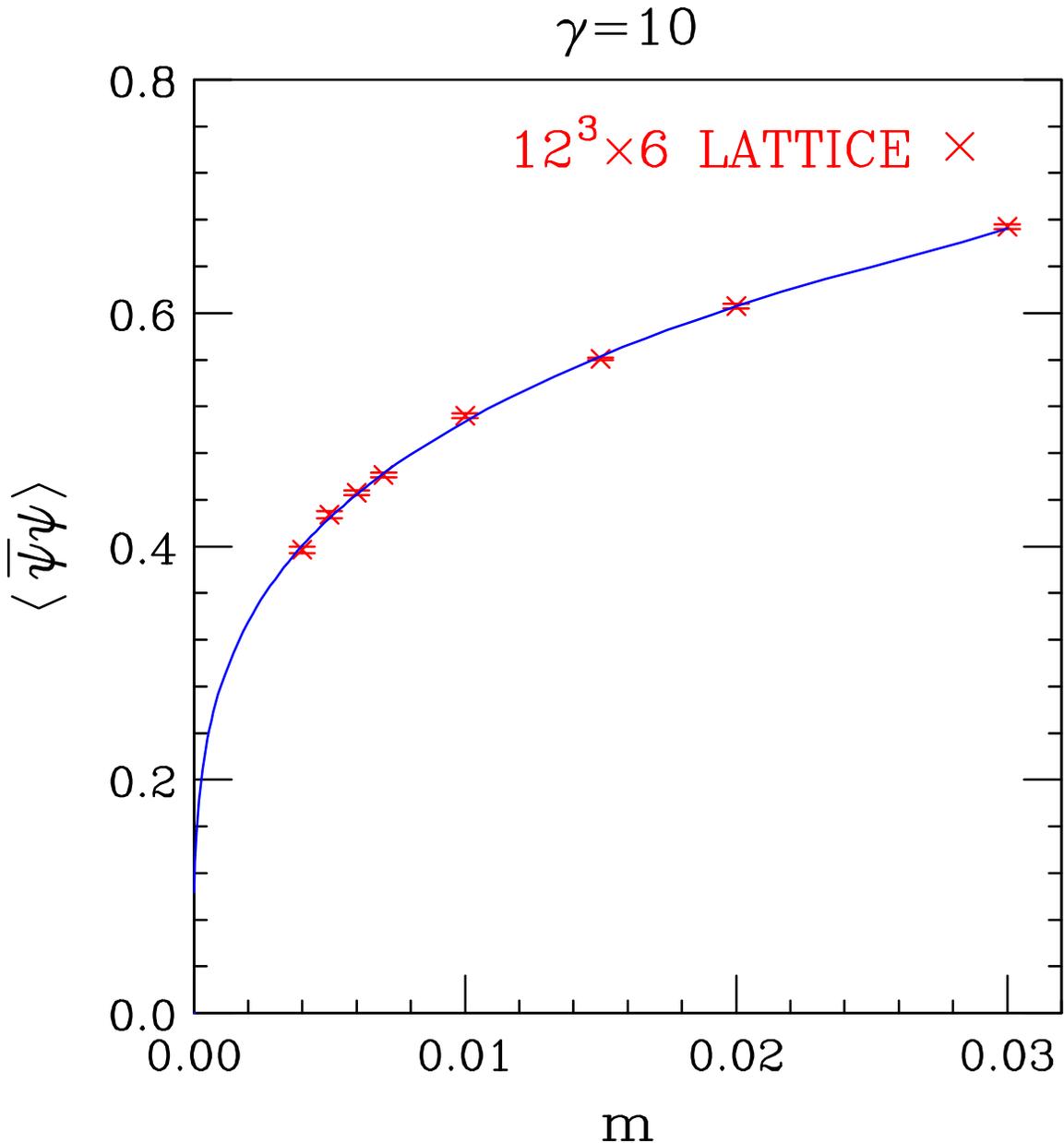}}
\caption{The order parameter plotted against $m$ at criticality on a 
$12^3 \times 6$ lattice. The solid line is the fit 
$\langle\bar{\psi}\psi\rangle = am^{1/ \delta}$. 
\label{fig:del612}}
\end{figure}

A second independent critical index is $\delta$ which controls the size of the
order parameter at the critical temperature as the quark mass is turned on
with $\beta$ fixed at $\beta_c$. The order parameter should scale as
$\langle\bar{\psi}\psi\rangle = am^{1/ \delta}$.  We accumulated data at quark
masses ranging from $0.004$ to $0.030$, as shown in Figure~\ref{fig:del612} on
a $12^3 \times 6$ lattice with the strength of the four fermi coupling set to
$\gamma = 10$. (We repeated this for $\gamma=20$.) Since the quark mass is
different from zero in these simulations, the real order parameters
$\langle\sigma\rangle$, $\langle\bar{\psi}\psi\rangle$ can be measured rather
than their absolute values. Our best fit gives $a=1.66(2)$ and critical index
$\delta=3.89(3)$. The fact that this fit has only a 4\% confidence is due to
the point at $m=0.01$ which lies 2.5 standard deviations above the curve and
is a problem for {\it any} smooth fit to the ``data''. Its removal increases
the confidence to 33\% with negligible change in the parameters of the fit.
This lies between 3 and 5, the $\delta$ values for the 2 leading symmetry
breaking operators at the tricritical point. Since 
$\langle\bar{\psi}\psi\rangle$ could be expected to be a linear combination
of these operators (and a third operator with $\delta=1$ which is normally
ignored), the more general scaling relation at $\beta=\beta_c$ is given by
\begin{equation}
\langle\bar{\psi}\psi\rangle = a\phi + \frac{1}{3}b\phi^3 + \frac{1}{5}c\phi^5
\end{equation}
with $\phi$ the solution of
\begin{equation}
\phi^5 - m ( a + b \phi^2 + c \phi^3) = 0
\end{equation}
A fit to this form over the whole mass range yields $a=1.05(2)$, $b=2.5(4)$
and $c=-3.0(1.4)$. Although the confidence level of this fit is only 2\%,
removing the $m=0.01$ point improves this to an acceptible 34\%.

\begin{figure}[htb]
\epsfxsize=6in
\centerline{\epsffile{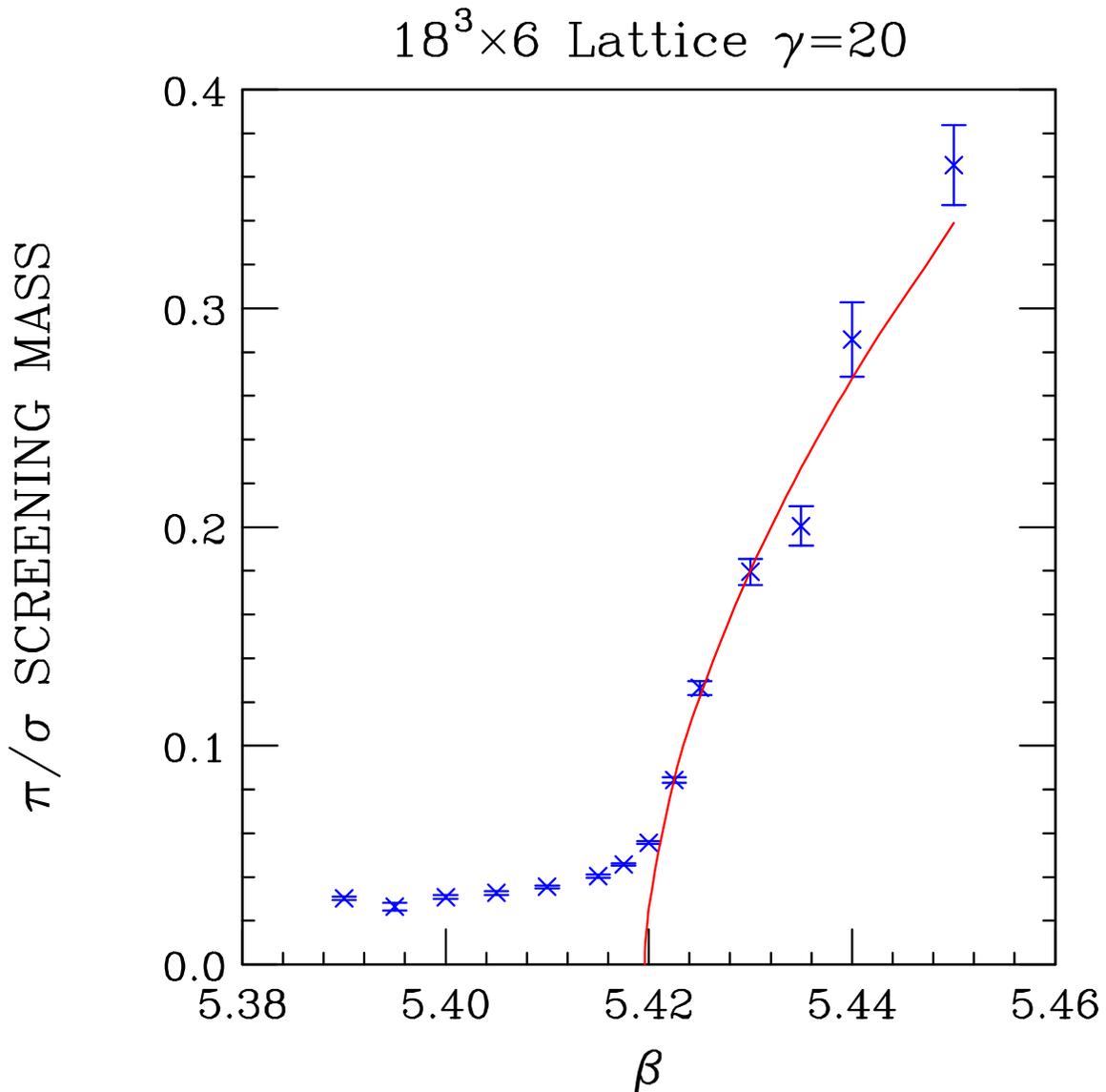}}
\caption{The pion screening mass plotted against $\beta$ on a 
$18^3 \times 6$ lattice. 
The solid line is the fit discussed in the
text.\label{fig:pi618}}
\end{figure}

The hadonic screening lengths determine the manner in which chiral symmetry is
restored as we pass through the transition from hadronic matter to a
quark-gluon plasma. Here we concentrate on the $\pi$ and $\sigma(f_0)$
propagators which can be calculated from the $\pi$ and $\sigma$ auxiliary
fields. For these we stored the $\pi$ and $\sigma$ fields averaged over each
z-slice of the lattice, every 2 time units. Because the orientation of the
condensate changes continuously throughout the runs, we relabeled the
component of these fields in the direction of the condensate as $\sigma$ and
the orthogonal component as $\pi$ for each configuration. These fields were
then correlated to calculate the average $\sigma$ and $\pi$ screening
propagators. The sum of these propagators is easily seen to give the joint
$\sigma/\pi$ propagator in the plasma phase. The screening mass obtained from
this unsubtracted propagator is plotted in Figure~\ref{fig:pi618} as a
function of $\beta$. It flattens out in the hadronic matter phase when the
fields develop a vacuum expectation value heralding a truly massless pion.
While good fits of the pion propagator to a massless boson propagator are
obtained in this region, fits in which the mass is allowed to float typically
give masses $\sim 0.2$. To get clearer evidence for a massless pion would
require a lattice of larger spatial extent (and/or better statistics). 
Until then we appeal to Goldstone's theorem for the knowledge that the pion is
exactly massless in the chirally broken phase \cite{LKS}. The fit for the
screening mass in the plasma phase is also shown. The fit used the data points
at $\beta = 5.4225$ through $\beta = 5.45$ and used the form $a(\beta -
\beta_c)^{\nu}$. It predicted a correlation length exponent of $\nu = 0.59(7)$, 
an amplitude $a = 2.66(71)$ and a critical coupling $\beta_c = 5.4196(7)$. As
is clear from the figure, the data here suffers from errors, both statistical
and systematic, that are larger than in our other analyses. The prediction
$\nu = 0.59(7)$ is compatible with the $0.5$ of the tricritical point.

In summary, our measurements of $\beta_{mag}$ at $N_t=6$, made possible by our 
new action which permits zero mass simulations, strongly favour tricritical
scaling at the chiral phase transition over the $O(4)/O(2)$ transition
expected from universality arguements. Combined with the first order transition
we have observed at $N_t=4$ it gives hope that $O(4)/O(2)$ scaling will be
observed for $N_t \ge 8$. This `anomalous' behaviour for $N_t=4,6$ offers a
potential explanation as to why simulations with the standard staggered fermion
action produce unexpected results \cite{milc,karsch,aoki}.

The extra symmetry breaking operator at the tricritical point provides a
possible explanation of the scaling of the chiral condensate with mass at
$\beta_c$ and the measured value of $\delta$. $\nu$ is consistent with its
tricritical value.

Simulations on $N_t=8$ lattices should be performed to determine whether the
continuum limit shows universal critical behaviour. Lattices with larger
spatial extent are needed to obtain better predictions of $\nu$.

\section{ACKNOWLEDGEMENTS}

These computations were performed on the CRAY C90/J90/SV1's and T3E at NERSC,
and on a T3E at SGI/Cray. We would like to thank J.-F.~Laga\"{e} who
contributed in the earlier stages of this work. We also thank Dr.~M.~Stephanov
for discussions of tricritical phenomena. This work was supported by the U.~S.
Department of Energy under contract W-31-109-ENG-38, and the National Science
Foundation under grant NSF-PHY96-05199.

\end{document}